\documentclass[aps,pre,onecolumn,nofootinbib]{revtex4}
\usepackage{graphicx}
\usepackage{dcolumn}
\usepackage{bm}
\usepackage{amsmath}
\usepackage{mathrsfs}
\usepackage{subfigure}

\usepackage[dvips]{color}

\begin{document}

\title{Revisiting finite size effect of percolation in degree correlated networks}

\author{Shogo Mizutaka}
\email{mizutaka@jaist.ac.jp}
\affiliation{Japan Advanced Institute of Science and Technology, 1-1 Asahidai, Nomi 923-1211, Japan}
\author{Takehisa Hasegawa}
\email{takehisa.hasegawa.sci@vc.ibaraki.ac.jp}
\affiliation{Department of Mathematics and Informatics, Ibaraki University, 2-1-1 Bunkyo, Mito 310-8512, Japan}

\date{\today}
\begin{abstract}
In this study, we investigate bond percolation in networks that have the Poisson degree distribution and a nearest-neighbor degree-degree correlation. Previous numerical studies on percolation critical behaviors of degree-correlated networks remain controversial. We perform finite-size scaling for the peak values of the second-largest cluster size and the mean cluster size and find a large finite-size effect when a network has a strong degree-degree correlation. Evaluating the size dependence of estimated critical exponents carefully, we demonstrate that the bond percolation in the networks exhibits the mean-field critical behavior, independent of the strength of their nearest-neighbor degree correlations.\end{abstract}
\maketitle
\section{INTRODUCTION}
Percolation in complex networks has garnered significant attention in various scientific fields since the beginning of network science \cite{li2021percolaiton,newman_text}.
In site (bond) percolation in a network, nodes (edges) are occupied with a probability of $p$ and unoccupied otherwise.
The network undergoes continuous transition at a percolation threshold of $p_{\rm c}$ above which a giant component emerges.
The relation between the critical exponents associated with the percolation transition and structures of underlying networks is an essential topic in statistical physics.
Percolation in Erd\H{o}s-R\'{e}nyi (ER) random networks that have a Poisson degree distribution is characterized by a standard mean-field class similar to lattices above the upper critical dimension.
In contrast, when networks have a scale-free property in which the degree distribution $P(k)$ decays according to a power-law, that is, $P(k)\sim k^{-\lambda}$, the heterogeneity of the degrees affects the percolation critical behavior \cite{cohen2002percolation}.
For random scale-free networks with $2<\lambda<4$, percolation critical exponents depend on the exponent $\lambda$, whereas for networks with $\lambda>4$, they are the same as those for ER random networks.

It has been discussed that critical behaviors in networks depend on degree-correlated structures
\cite{Goltsev08,mizutaka2020percolation,Noh07,Valdez11,schmeltzer2014percolation}.
The nearest-neighbor degree correlation, which is a correlation between degrees of directly connected nodes by edges, is considered a first step in characterizing degree-correlated structures \cite{orsini2015quantifying}.
The Pearson's correlation coefficient of nearest degrees, namely, the assortativity coefficient, quantifies the nearest-neighbor degree correlation of a network as follows:
\begin{align}
r = \frac{4 \langle k k' \rangle_e - \langle k+k' \rangle_e^2}{2 \langle k^2+k'^2 \rangle_e - \langle k+k' \rangle_e^2},	
\end{align}
where $\langle f(k,k')\rangle_e$ denotes the average of $f(k,k')$ over the joint probability $P(k,k')$ that two ends of a randomly selected edge are degree-$k$ and -$k'$ nodes.
Networks with $r>0$ $(r<0)$ are assortative (disassortative) in which similar (dissimilar) degree nodes will likely be connected to each other.
The percolation threshold is lowered (raised) by the assortative (disassortative) mixing compared with that of uncorrelated networks \cite{Newman02,Newman03,Goltsev08}.
Moreover, theoretical studies revealed that nearest-neighbor degree correlations change critical behaviors of the percolation transition in some cases \cite{Goltsev08,mizutaka2020percolation}.

In contrast to theoretical studies, however, numerical studies on critical behaviors of degree correlated networks still remain controversial.
Noh investigated bond percolation in degree-correlated ER networks generated by an exponential random graph (ERG) model\footnote{In the ERG model, we consider a Hamiltonian $H(G)=-J\sum_{i>j} k_i k_jA_{ij}$ of a network $G$, where $A_{ij}$ denotes the adjacency matrix and $J$ is a control parameter for the correlation. The following two steps are repeated until the steady state is achieved: (i) two edges are selected randomly from a network $G$, and (ii) rewiring from $G$ to $G'$ is accepted with probability ${\rm min}[1,\exp(-H(G')+H(G))]$ and the network renames $G$. See \cite{Noh07} for details.}, by finite-size scaling \cite{Noh07}.
\begin{figure*}[t]
\begin{center}
\includegraphics[width=1\textwidth]{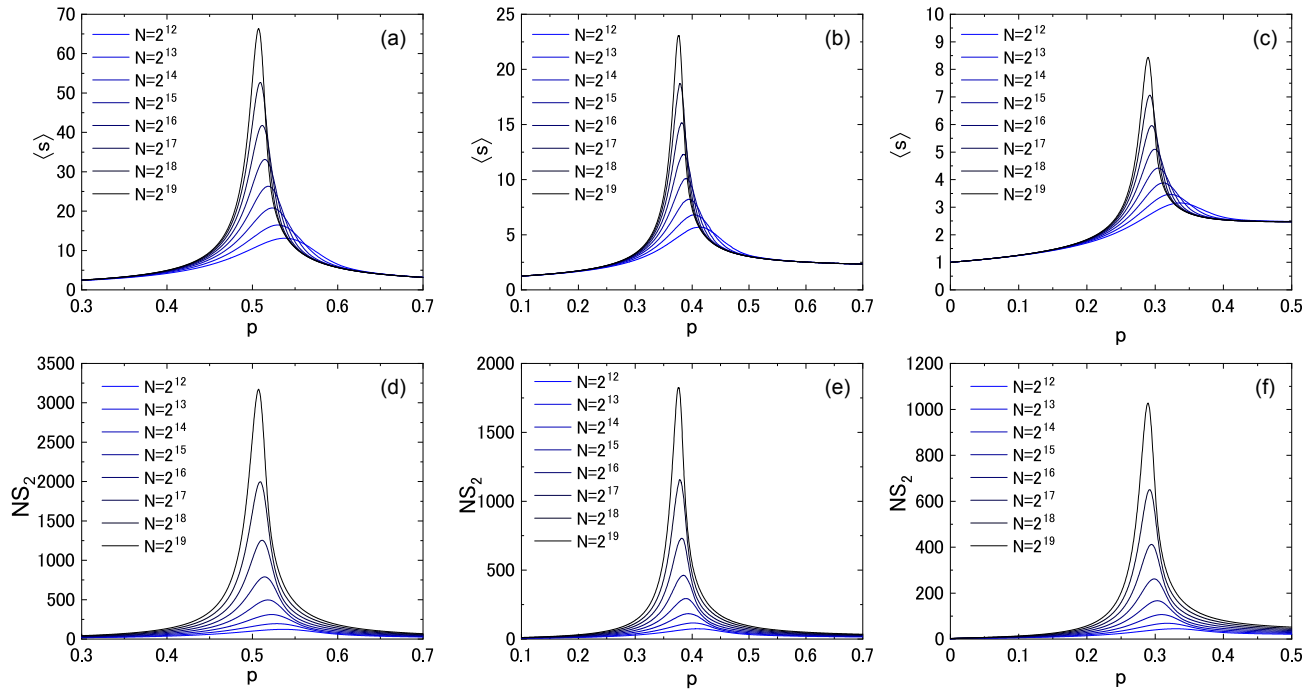}
\caption{
Occupation probability $p$ dependences of (a)--(c) average cluster size $\langle s\rangle$ and (d)--(f) size $NS_2$ of the second largest cluster. Panels in each row from left to right represent the results for (un-)correlated ER networks with $r=0$, $r=0.5$, and $r=0.8$, respectively. These are obtained for networks with each size averaged over $10^3$ samples.
}
\label{fig:sav_S2}
\end{center}
\end{figure*}
The non-diverging mean size of finite clusters at the percolation threshold is observed for the assortative ER networks, implying that the type of transition differs from that in uncorrelated ER networks.
Valdez {\it et al}.\  introduced a different algorithm, called the local optimal algorithm (LOA), to incorporate a degree correlation in ER networks \cite{Valdez11}.
In the LOA, the following rewiring process is considered: (i) two edges are selected randomly from a network, and (ii) the configuration that increases (decreases) most the assortativity coefficient $r$ is employed as a rewiring to positively (negatively) correlate.
The rewiring process continues until $r$ reaches the desired value.
They compared bond percolation on assortative ER networks generated by LOA with those generated by the ERG model in the same value of assortativity coefficient $r$.
In the bond percolation transition of LOA-based assortative ER networks, the mean size of finite clusters diverges in power-law with the system size, and a non-mean-field exponent characterizes the divergence.
Thus, the percolation critical behavior for LOA-based assortative networks differs from those of either mean-field type or assortative networks generated by the ERG algorithm.
They also reported that the LOA and ERG model generate different correlated structures in a long-range even though the assortativity coefficient $r$ coincides with two models, and further speculated that long-range correlation could change drastically critical behavior on networks with the same value of assortativity coefficient $r$.
It has not been investigated whether percolation critical behaviors on these networks would change if long-range degree correlations are broken (and only nearest-neighbor degree correlations are retained).
Long-range degree correlations would be broken by edge randomization preserving the joint probability $P(k,k')$.
In this study, we investigate bond percolation on the LOA-based networks with the edge randomization preserving the joint probability $P(k,k')$.
First, we generate LOA-based networks with a value of $r$ and rewire those networks by the edge randomization retaining $P(k,k')$.
Next, bond percolation processes are examined on randomized-LOA-based ER networks with various values of $r$.
From extensive finite-size scaling, we demonstrate that the critical behaviors of bond percolation on randomized-LOA-based networks with any $r$ values are of the standard mean-field.

\section{Methods}\label{sec:methods}
\subsection{Correlated Erd\H{o}s-R\'{e}nyi networks}
\label{sec:correlated_networks}
We introduce the preparation of degree-correlated networks that are treated in this study.
First, we generate an ER network with a network size of $N$. The degree distribution is $P(k)=\langle k\rangle^k e^{-\langle k\rangle}/k!$, where $\langle k\rangle$ denotes the average degree.
In this study, we set the average degree as $\langle k\rangle=2$ in all simulations.
Next, we rewire edges of the network based on the LOA until the assortativity coefficient $r$ reaches the desired value.
Finally, we break a long-range degree correlation in the LOA-based network by the subsequent randomization.
Two randomly selected edges are swapped if the degree of either end of an edge matches the degree of either end of the other edge. The randomization is repeated a large number of times.
A generated network has a nearest-neighbor degree correlation being maximally randomized with a fixed $P(k,k')$, and the assortativity coefficient $r$ is maintained constant.
All networks in this study are generated from the aforementioned algorithm except for ERG-based networks in Fig.~\ref{fig:LOA_ERG_rl}.
\begin{figure*}[t]
\begin{center}
\includegraphics[width=1\textwidth]{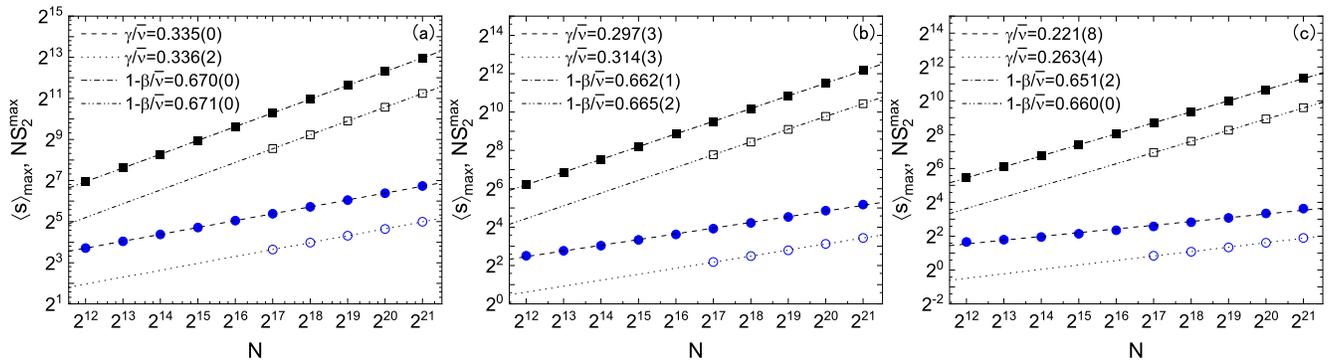}
\caption{
Scalings for peak values $\langle s\rangle_{\rm max}$ and $NS_{2}^{\rm max}$. Panels (a), (b), and (c) represent results for the assortativity coefficient $r=0$, $0.5$, and $0.8$, respectively.  
}
\label{fig:scaling}
\end{center}
\end{figure*}

\subsection{Finite-size scaling analysis}\label{sec:FSS}
The present study concentrates on the bond percolation process where each edge is occupied with the probability $p$ and unoccupied otherwise.
We measure the average size $\langle s\rangle$ of finite clusters and the relative size $S_{2}$ of the second largest cluster to estimate associated critical exponents that are introduced below.
Figure~\ref{fig:sav_S2} shows $\langle s\rangle$ [panels (a), (b), and (c)] and $NS_{2}$ [panels (d), (e), and (f)] as functions of the occupation probability $p$.
In panels (a), (b), and (c) of Fig.~\ref{fig:sav_S2}, which plots the results for uncorrelated ($r=0$), $r=0.5$, and $r=0.8$, respectively, we find peaks that become sharp with increasing $N$ reported as ~\cite{Valdez11}. We observe the same fashion in $NS_{2}$ in panels (d), (e), and (f).
In addition, as observed in Fig.~\ref{fig:sav_S2}, both peaks of $\langle s\rangle$ and $NS_{2}$ are suppressed as increasing the assortativity coefficient $r$.

Applying a finite-size scaling analysis to the percolation transition in the correlated ER networks, we discuss the percolation critical behavior.
We assume $\langle s\rangle$ has a scaling form as follows:
\begin{equation}
	\langle s\rangle(p, N) \sim N^{\gamma/\bar{\nu}}f_{\langle s\rangle}[(p-p_{\rm c})N^{1/\bar{\nu}}],
	\label{eq:ave_cluster_size}
\end{equation}
where $\gamma$ denotes a critical exponent associated with $\langle s\rangle \sim |p-p_{\rm c}|^{-\gamma}$, $\bar{\nu}$ is a critical exponent in a finite-size scaling, and $f_{\langle s\rangle}(x)$ is the scaling function of $\langle s\rangle$.
Similarly, the size $S_{2}$ of the second largest cluster behaves as follows:
\begin{equation}
	S_{2}(p, N) \sim N^{-\beta/\bar{\nu}}g_{S_{2}}[(p-p_{\rm c})N^{1/\bar{\nu}}],
	\label{eq:S2}
\end{equation}
where $\beta$ denotes a critical exponent associated with the relative size of clusters, that is, the order parameter \cite{almeira2020scaling,zhu2017finite} and $g_{S_{2}}(x)$ is the scaling function for $S_{2}$.
Both values $\langle s\rangle$ and $S_{2}$ have a peak and their peak positions $p_{\rm max}$ scale as follows:
\begin{equation}
	p_{\rm max}(N)=p_{\rm c}+a_{\langle s\rangle(S_{2})}N^{-1/\bar{\nu}},
	\label{eq:peak_position}
\end{equation}
where $a_{\langle s\rangle}$ and $a_{S_{2}}$ are constants.
Substituting Eq.~(\ref{eq:peak_position}) in Eqs.~(\ref{eq:ave_cluster_size}) and (\ref{eq:S2}), we obtain the following simple scaling forms:
\begin{align}
	\langle s\rangle_{\rm max}&\sim N^{\gamma/\bar{\nu}} \label{eq:for_gamma}\\
	NS_{2}^{\rm max}&\sim N^{1-\beta/\bar{\nu}}. \label{eq:for_beta}
\end{align}
Here, $\langle s\rangle_{\rm max}=\langle s\rangle(p_{\rm max}, N)$ and $S_{2}^{\rm max}=S_{2}(p_{\rm max}, N)$.
The ratios $\beta/\bar{\nu}$ and $\gamma/\bar{\nu}$ of critical exponents satisfy a known relation in continuous phase transition \cite{Stauffer}:
 \begin{equation}
 	2\frac{\beta}{\bar{\nu}}+\frac{\gamma}{\bar{\nu}}=1.
 	\label{eq:scaling_relation}
 \end{equation}
We can use the relation (\ref{eq:scaling_relation}) to verify whether critical exponents are sufficiently estimated by finite-size scaling\footnote{Notably, the relation (\ref{eq:scaling_relation}) is a necessary condition.}.
By estimating exponents $\gamma/\bar{\nu}$ and $1-\beta/\bar{\nu}$ through (\ref{eq:for_gamma}) and (\ref{eq:for_beta}) by numerical simulations and testing the relation (\ref{eq:scaling_relation}), we investigate the critical behavior of percolation transition in correlated ER networks in the next section.

\section{results}\label{sec:results}
Figure~\ref{fig:scaling} shows scalings in Eqs.~(\ref{eq:for_gamma}) and (\ref{eq:for_beta}) for several values of assortativity coefficient $r$.
To obtain a peak value (each symbol in Fig.~\ref{fig:scaling}), we performed a Gaussian fit for the top $5\%$ of the data from the maximum value.
In each panel, we show $\langle s\rangle_{\rm max}$ (filled blue circles) and $NS_{2}^{\rm max}$ (filled black squares) for networks with $N=2^{12}$, $2^{13}$, ..., and $2^{21}$, and vertically shifted $\langle s\rangle_{\rm max}$ (open blue circles) and $NS_{2}^{\rm max}$ (open black squares) for networks of the top five sizes ($N=2^{17}$, $2^{18}$, $2^{19}$, $2^{20}$, and $2^{21}$) to compare estimated slopes.
In all panels of Fig.~\ref{fig:scaling}, linear relations of the plots on the log--log scale can be confirmed.
As observed from panel (a) in Fig.~\ref{fig:scaling}, which displays the results for uncorrelated ER networks ($r=0$),
estimates of $\gamma/\bar{\nu}$ (slope of dashed line) and $1-\beta/\bar{\nu}$ (slope of dashed-dotted line) are consistent with mean-field exponents ($\gamma_{\rm MF}/\bar{\nu}_{\rm MF}=1/3$ and $\beta_{\rm MF}/\bar{\nu}_{\rm MF}=1/3$), and are identical to those estimated from the top five sizes (slopes of dotted and dashed-double-dotted lines).
Here, $\gamma_{\rm MF}$, $\beta_{\rm MF}$, and $\bar{\nu}_{\rm MF}$ represent the mean-field exponents.
This implies that the finite size effect for the percolation in uncorrelated ER networks is sufficiently small even though the network size is relatively small such as $N=O(10^4)$.
Figure~\ref{fig:scaling} (b) shows that for $r=0.5$, dashed-dotted and dashed-double-dotted lines associated with $\beta/\bar{\nu}$ are parallel and their slopes are consistent with the mean-field value $\beta_{\rm MF}/\bar{\nu}_{\rm MF}=1/3$, whereas dashed and dotted lines associated with $\gamma/\bar{\nu}$ are not parallel.
In addition, estimates of $\gamma/\bar{\nu}$ and $\beta/\bar{\nu}$ from all data ($N=2^{12}$, $2^{13}$, ..., $2^{21}$) do not correspond to those estimated from the top five sizes of data for strongly assortative ER networks with $r=0.8$ [see lines in Fig.~\ref{fig:scaling} (c)].
These discords are caused by the finite size effect for the bond percolation in correlated ER networks.

\begin{figure}[t!]
\begin{center}
\includegraphics[width=0.45\textwidth]{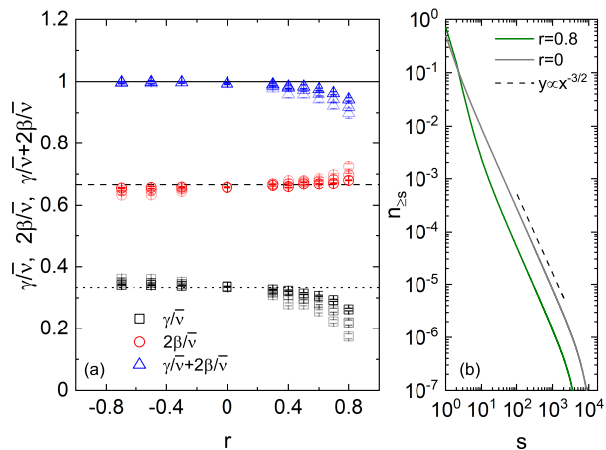}
\caption{(a) Estimates of $\gamma/\bar{\nu}$ (black squares), $2\beta/\bar{\nu}$ (red circles), and their sum (blue triangles) as a function of assortativity coefficient $r$. The light, medium, dark squares (circles) are estimated from peak values $\langle s\rangle_{\rm max}$ ($S_{2}^{\rm max}$) with $N=2^{12},2^{13},2^{14},2^{15},2^{16}$, $N=2^{12},2^{13},\cdots,2^{21}$, and $N=,2^{17},2^{18},2^{19},2^{20},2^{21}$, respectively. Dotted and dashed lines represent the mean-field values $\gamma_{\rm MF}/\bar{\nu}_{\rm MF}$ and $\beta_{\rm MF}/\bar{\nu}_{\rm MF}$, respectively. Solid line indicates the scaling relation (\ref{eq:scaling_relation}).
(b) Cumulative distribution of the size of finite clusters at the percolation threshold $p_{\rm max}(N)$. The green and grey lines represent the results for ER networks with $r=0.8$ and $r=0.0$, respectively. These were obtained from $10^2$ samples. We run the bond percolation process $10^2$ times in each sample. The network size is $N=2^{21}$ nodes. Estimating the slopes of green and grey lines within $[10^2,10^3]$, we obtained $1.55$ and $1.53$, respectively.}
\label{fig:r_exponents}
\end{center}
\end{figure}
In Fig.~\ref{fig:r_exponents} (a), we depict the assortativity coefficient $r$ dependences of estimated critical exponents.
The black squares and red circles represent estimates of $\gamma/\bar{\nu}$ and $2\beta/\bar{\nu}$, respectively.
Dotted and dashed horizontal lines in Fig.~\ref{fig:r_exponents} (a) represent mean-field values $\gamma_{\rm MF}/\bar{\nu}_{\rm MF}=1/3$ and $2\beta_{\rm MF}/\bar{\nu}_{\rm MF}=2/3$, respectively.
The blue triangles are the sum of estimates of $\gamma/\bar{\nu}$ and $2\beta/\bar{\nu}$.
We draw the solid line that represents a scaling relation (\ref{eq:scaling_relation}).
The difference of color depth in each symbol type corresponds to the data difference used for the finite-size scaling analysis.
The light, medium, dark colors represent the values estimated from networks with $N=2^{12},2^{13},2^{14},2^{15},2^{16}$, $N=2^{12},2^{13},\cdots,2^{21}$, and $N=,2^{17},2^{18},2^{19},2^{20},2^{21}$, respectively.
For uncorrelated networks, circles and squares with each color depth overlap on the dashed and dotted lines, respectively.
Thus, the estimates do not strongly depend on the data used for scalings, implying that the finite-size effect is weak, as shown in Fig.~\ref{fig:scaling}.
In contrast, there exist finite-size effects in disassortative and assortative cases.
In red circles for each value of $r$, the darker the color depth, the more the estimates of $\beta/\bar{\nu}$ approach the dashed line, implying that $\beta/\bar{\nu}=\beta_{\rm MF}/\bar{\nu}_{\rm MF}$.
The result of $S_{2}^{\rm max}$ indicates that the bond percolation in correlated ER networks with any of $r$ is of mean-field class.
For $r<0.5$, estimates of $\gamma/\bar{\nu}$ are also consistent with $\gamma_{\rm MF}/\bar{\nu}_{\rm MF}$ and the scaling relation (\ref{eq:scaling_relation}) is valid, inevitably.
For strongly assortative cases ($r\ge 0.5$), in contrast, we can confirm a large finite-size effect in scaling for $\langle s\rangle_{\rm max}$.
The differences in $\gamma/\bar{\nu}$ values estimated from three data sets are large compared with those for $r< 0.5$, implying that the estimated values of $\gamma/\bar{\nu}$ for $r\ge 0.5$ do not converge sufficiently.

Figure~\ref{fig:r_exponents} (b) shows the cumulative distribution $n_{\ge s}$ of finite size clusters at the percolation threshold $p_{\rm max}(N)$.
The green and grey lines represent the results for $r=0.8$ and uncorrelated networks, respectively.
Although the grey line is linear in the log--log plot in the entire region of $s$, the green line deviates from the straight line in the small $s$ region.
In strongly assortative ER networks, there are several small clusters contributing to assortative mixing. 
Such disconnected small clusters are not involved with the criticality of percolation transition and suppress the mean cluster size $\langle s\rangle_{\rm max}$, resulting in a strong finite-size effect in $\langle s\rangle_{\rm max}$.
At the percolation threshold, the distribution $n_{\ge s}$ is characterized by a power-law fashion, that is, $n_{\ge s}\sim s^{-\tau+1}$, where $\tau$ is a critical exponent.
Estimated exponents $\tau$ values for correlated ER networks (green line) and uncorrelated networks (grey line) are obtained by $1.55$ and $1.53$, respectively.
These are comparable with the mean-field value $\tau_{\rm MF}=5/2$, which supports that percolation in strongly assortative networks is of mean-field class.

\section{Conclusion \& Discussion}\label{sec:conclusion}
We investigated the bond percolation in ER networks with (only) nearest-neighbor degree-degree correlations, which are generated by the LOA and maximally randomized while retaining the joint probability $P(k,k')$.
By performing finite-size scaling for the peak values of size of the second largest cluster and of the mean cluster size, $NS_{2}^{\rm max}$ and $\langle s\rangle_{\rm max}$, we estimated critical exponents $\beta/\bar{\nu}$ and $\gamma/\bar{\nu}$.
The estimates of $\beta/\bar{\nu}$ converge to the mean-field value for all assortative and disassortative ER networks, implying that the percolation transition of correlated ER networks is the mean-field.
With regard to $\gamma/\bar{\nu}$, estimated values are consistent with the mean-field value for weakly assortative and all disassortative ER networks, although a large finite-size effect is observed in scaling for $\langle s\rangle_{\rm max}$ in strongly assortative cases.
Even if the large finite-size effect does not provide satisfactory convergence estimates of $\gamma/\bar{\nu}$, we confirmed that the exponent $\tau$ associated with the cluster size distribution is comparable with the mean-field value in strongly assortative networks.
Overall, our simulations indicate that the bond percolation in correlated ER networks exhibits the mean-field critical behavior, independent of the strength of their nearest-neighbor degree correlations.
\begin{figure}[t!]
\begin{center}
\includegraphics[width=0.45\textwidth]{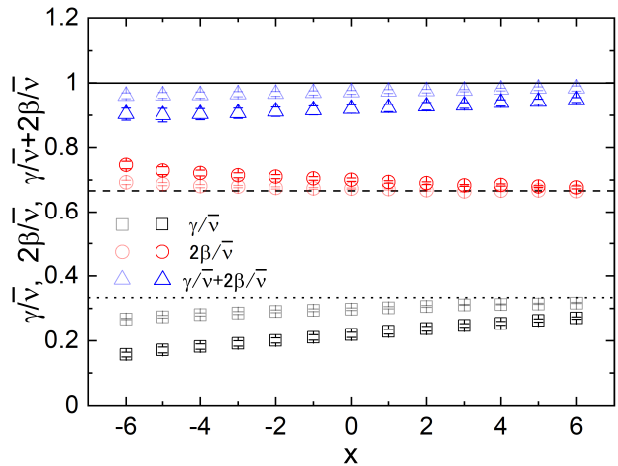}
\caption{Effect of the network sizes of data used for scaling on estimated critical exponents $\gamma/\bar{\nu}$ and $2\beta/\bar{\nu}$ for LOA-based degree correlated ER networks.
The lighter and darker symbols are results for $r=0.5$ and $r=0.8$, respectively.
Values estimated by scalings for the results with $N=2^{12}$, $2^{13}$, ..., $2^{21}$ are located at $x=0$. At $x=a(x=-a)$, values estimated from the data are thinned out by $a$ from $N=2^{12}$ ($N=2^{21}$). For instance, points at $x=5$ are estimated from the data for $N=2^{17}$, $2^{18}$, $2^{19}$, $2^{20}$, and $2^{21}$. Thus, the larger $x$ is, the more we estimate exponents from large network sizes only. 
}
\label{fig:data_dependence_LO}
\end{center}
\end{figure}

\begin{figure}[t!]
\begin{center}
\includegraphics[width=0.45\textwidth]{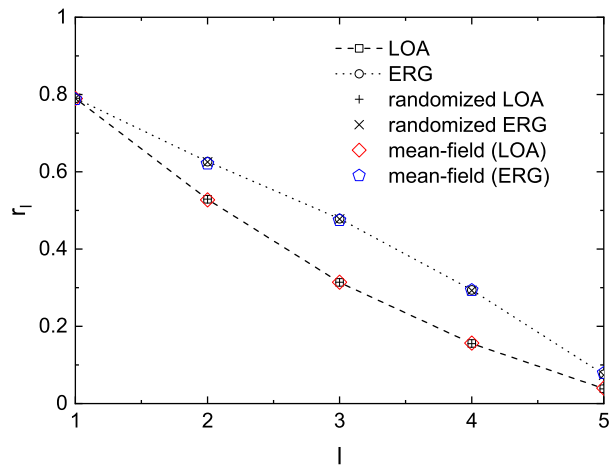}
\caption{Distance $l$ dependence of $r_l$ values for LOA- and ERG-based ER networks (squares and circles), randomized-LOA-based and randomized-ERG-based ER networks (pluses and crosses), and mean-field results predicted by the numerically obtained joint probability $P(k,k')$ in LOA and ERG (red diamonds and blue pentagons). We employ ER networks with $\langle k\rangle=2$, $r=0.79$, and $N=2^{15}$. The simulation was performed over $10^3$ realizations. In the ERG algorithm, $J=1$ was employed and each network generation terminated at $10^3$ Monte-Carlo steps.}
\label{fig:LOA_ERG_rl}
\end{center}
\end{figure}
In ~\cite{Valdez11}, the bond percolation in correlated networks generated by the LOA was treated.
The authors performed finite-size scaling for the peak value of the mean cluster size by using relatively small network sizes and reported the estimated value as $\gamma/\bar{\nu}= 0.16$.
To test a finite-size effect of percolation in LOA-based correlated ER networks, we plotted the size dependence of estimated values of $\gamma/\bar{\nu}$, $\beta/\bar{\nu}$ and the scaling relation (\ref{eq:scaling_relation}) in Fig.~\ref{fig:data_dependence_LO}.
In this figure, the larger $x$ is, the more estimated values are provided by scaling for large network sizes only.
The estimate of $\gamma/\bar{\nu}$ ($\beta/\bar{\nu}$) increases (decreases) monotonically with increasing $x$, implying that a large finite-size effect exists in the percolation of LOA-based correlated networks.
The value of $\gamma/\bar{\nu}$ for a large $x$ is closer to the mean-field value than that reported in ~\cite{Valdez11}.
In addition, the value of $\beta/\bar{\nu}$ converges to the mean-field value.
These results imply that the percolation transition in LOA-based ER networks is of the mean-filed.

Does the edge randomization retaining the joint probability $P(k,k')$ break the long-range degree correlation in LOA- and ERG-based networks? For both algorithms, we compared the following three $r_l$ values that are the Pearson's coefficient between degrees of nodes separated by $l$: (i) original $r_l$ for the networks generated by the LOA (ERG), (ii) $r_l$ for LOA-based (ERG-based) networks while preserving $P(k,k')$, and (iii) $r_l$ analytically estimated from the mean-field theory \cite{fujiki2020identification} using only numerically obtained $P(k,k')$ values from LOA-based (ERG-based) networks (See \cite{Valdez11,mayo2015long,fujiki2018general} for the details of $r_l$).
For $l=1$, $r_l$ coincides with the assortativity coefficient $r$.
As shown in Fig.~\ref{fig:LOA_ERG_rl}, we can confirm that all $r_l$ values are the same in both cases of LOA and ERG, implying that the difference in $r_l$ at $l\ge 2$ between LOA and ERG is not as a result of non-trivial long-range degree correlations, but only the difference in $P(k,k')$.
Although the percolation transition is of mean-field in LOA-based assortative networks, it is of infinite-order in ERG-based assortative networks.
This difference is attributed to only the difference of their joint probability $P(k,k')$. This issue should be studied further for better understanding of the relation between percolation transitions and degree-correlated structures of underlying networks.

\begin{acknowledgements}
S.M.\ was supported by JSPS (Japan) KAKENHI (Grant Number 21K13853).
T.H.\ acknowledges the financial support from JSPS (Japan) KAKENHI (Grant Number JP19K03648).
\end{acknowledgements}

\bibliography{ref.bib}

\begin{thebibliography}{17}
\expandafter\ifx\csname natexlab\endcsname\relax\def\natexlab#1{#1}\fi
\expandafter\ifx\csname bibnamefont\endcsname\relax
  \def\bibnamefont#1{#1}\fi
\expandafter\ifx\csname bibfnamefont\endcsname\relax
  \def\bibfnamefont#1{#1}\fi
\expandafter\ifx\csname citenamefont\endcsname\relax
  \def\citenamefont#1{#1}\fi
\expandafter\ifx\csname url\endcsname\relax
  \def\url#1{\texttt{#1}}\fi
\expandafter\ifx\csname urlprefix\endcsname\relax\def\urlprefix{URL }\fi
\providecommand{\bibinfo}[2]{#2}
\providecommand{\eprint}[2][]{\url{#2}}

\bibitem[{\citenamefont{Li et~al.}(2021)\citenamefont{Li, Liu, Lü, Hu, Xu, and
  Zhang}}]{li2021percolaiton}
\bibinfo{author}{\bibfnamefont{M.}~\bibnamefont{Li}},
  \bibinfo{author}{\bibfnamefont{R.-R.} \bibnamefont{Liu}},
  \bibinfo{author}{\bibfnamefont{L.}~\bibnamefont{Lü}},
  \bibinfo{author}{\bibfnamefont{M.-B.} \bibnamefont{Hu}},
  \bibinfo{author}{\bibfnamefont{S.}~\bibnamefont{Xu}}, \bibnamefont{and}
  \bibinfo{author}{\bibfnamefont{Y.-C.} \bibnamefont{Zhang}},
  \bibinfo{journal}{Physics Reports} \textbf{\bibinfo{volume}{907}},
  \bibinfo{pages}{1} (\bibinfo{year}{2021}), ISSN \bibinfo{issn}{0370-1573},
  \bibinfo{note}{percolation on complex networks: Theory and application}.

\bibitem[{\citenamefont{Newman}(2010)}]{newman_text}
\bibinfo{author}{\bibfnamefont{M.~E.~J.} \bibnamefont{Newman}},
  \emph{\bibinfo{title}{Networks: an introduction}} (\bibinfo{publisher}{Oxford
  university press}, \bibinfo{year}{2010}).

\bibitem[{\citenamefont{Cohen et~al.}(2002)\citenamefont{Cohen, Ben-Avraham,
  and Havlin}}]{cohen2002percolation}
\bibinfo{author}{\bibfnamefont{R.}~\bibnamefont{Cohen}},
  \bibinfo{author}{\bibfnamefont{D.}~\bibnamefont{Ben-Avraham}},
  \bibnamefont{and} \bibinfo{author}{\bibfnamefont{S.}~\bibnamefont{Havlin}},
  \bibinfo{journal}{Physical Review E} \textbf{\bibinfo{volume}{66}},
  \bibinfo{pages}{036113} (\bibinfo{year}{2002}).

\bibitem[{\citenamefont{Goltsev et~al.}(2008)\citenamefont{Goltsev,
  Dorogovtsev, and Mendes}}]{Goltsev08}
\bibinfo{author}{\bibfnamefont{A.~V.} \bibnamefont{Goltsev}},
  \bibinfo{author}{\bibfnamefont{S.~N.} \bibnamefont{Dorogovtsev}},
  \bibnamefont{and} \bibinfo{author}{\bibfnamefont{J.~F.~F.}
  \bibnamefont{Mendes}}, \bibinfo{journal}{Phys. Rev. E}
  \textbf{\bibinfo{volume}{78}}, \bibinfo{pages}{051105}
  (\bibinfo{year}{2008}).

\bibitem[{\citenamefont{Mizutaka and Hasegawa}(2020)}]{mizutaka2020percolation}
\bibinfo{author}{\bibfnamefont{S.}~\bibnamefont{Mizutaka}} \bibnamefont{and}
  \bibinfo{author}{\bibfnamefont{T.}~\bibnamefont{Hasegawa}},
  \bibinfo{journal}{EPL (Europhysics Letters)} \textbf{\bibinfo{volume}{128}},
  \bibinfo{pages}{46003} (\bibinfo{year}{2020}).

\bibitem[{\citenamefont{Noh}(2007)}]{Noh07}
\bibinfo{author}{\bibfnamefont{J.~D.} \bibnamefont{Noh}},
  \bibinfo{journal}{Phys. Rev. E} \textbf{\bibinfo{volume}{76}},
  \bibinfo{pages}{026116} (\bibinfo{year}{2007}).

\bibitem[{\citenamefont{Valdez et~al.}(2011)\citenamefont{Valdez, Buono,
  Braunstein, and Macri}}]{Valdez11}
\bibinfo{author}{\bibfnamefont{L.~D.} \bibnamefont{Valdez}},
  \bibinfo{author}{\bibfnamefont{C.}~\bibnamefont{Buono}},
  \bibinfo{author}{\bibfnamefont{L.~A.} \bibnamefont{Braunstein}},
  \bibnamefont{and} \bibinfo{author}{\bibfnamefont{P.~A.} \bibnamefont{Macri}},
  \bibinfo{journal}{{EPL} (Europhysics Letters)} \textbf{\bibinfo{volume}{96}},
  \bibinfo{pages}{38001} (\bibinfo{year}{2011}).

\bibitem[{\citenamefont{Schmeltzer et~al.}(2014)\citenamefont{Schmeltzer,
  Soriano, Sokolov, and R{\"u}diger}}]{schmeltzer2014percolation}
\bibinfo{author}{\bibfnamefont{C.}~\bibnamefont{Schmeltzer}},
  \bibinfo{author}{\bibfnamefont{J.}~\bibnamefont{Soriano}},
  \bibinfo{author}{\bibfnamefont{I.~M.} \bibnamefont{Sokolov}},
  \bibnamefont{and}
  \bibinfo{author}{\bibfnamefont{S.}~\bibnamefont{R{\"u}diger}},
  \bibinfo{journal}{Physical Review E} \textbf{\bibinfo{volume}{89}},
  \bibinfo{pages}{012116} (\bibinfo{year}{2014}).

\bibitem[{\citenamefont{Orsini et~al.}(2015)\citenamefont{Orsini, Dankulov,
  Colomer-de Sim{\'o}n, Jamakovic, Mahadevan, Vahdat, Bassler, Toroczkai,
  Bogun{\'a}, Caldarelli et~al.}}]{orsini2015quantifying}
\bibinfo{author}{\bibfnamefont{C.}~\bibnamefont{Orsini}},
  \bibinfo{author}{\bibfnamefont{M.~M.} \bibnamefont{Dankulov}},
  \bibinfo{author}{\bibfnamefont{P.}~\bibnamefont{Colomer-de Sim{\'o}n}},
  \bibinfo{author}{\bibfnamefont{A.}~\bibnamefont{Jamakovic}},
  \bibinfo{author}{\bibfnamefont{P.}~\bibnamefont{Mahadevan}},
  \bibinfo{author}{\bibfnamefont{A.}~\bibnamefont{Vahdat}},
  \bibinfo{author}{\bibfnamefont{K.~E.} \bibnamefont{Bassler}},
  \bibinfo{author}{\bibfnamefont{Z.}~\bibnamefont{Toroczkai}},
  \bibinfo{author}{\bibfnamefont{M.}~\bibnamefont{Bogun{\'a}}},
  \bibinfo{author}{\bibfnamefont{G.}~\bibnamefont{Caldarelli}},
  \bibnamefont{et~al.}, \bibinfo{journal}{Nat. Commun.}
  \textbf{\bibinfo{volume}{6}}, \bibinfo{pages}{8627} (\bibinfo{year}{2015}).

\bibitem[{\citenamefont{Newman}(2002)}]{Newman02}
\bibinfo{author}{\bibfnamefont{M.~E.~J.} \bibnamefont{Newman}},
  \bibinfo{journal}{Phys. Rev. Lett.} \textbf{\bibinfo{volume}{89}},
  \bibinfo{pages}{208701} (\bibinfo{year}{2002}).

\bibitem[{\citenamefont{Newman}(2003)}]{Newman03}
\bibinfo{author}{\bibfnamefont{M.~E.~J.} \bibnamefont{Newman}},
  \bibinfo{journal}{Phys. Rev. E} \textbf{\bibinfo{volume}{67}},
  \bibinfo{pages}{026126} (\bibinfo{year}{2003}).

\bibitem[{\citenamefont{Almeira et~al.}(2020)\citenamefont{Almeira, Billoni,
  and Perotti}}]{almeira2020scaling}
\bibinfo{author}{\bibfnamefont{N.}~\bibnamefont{Almeira}},
  \bibinfo{author}{\bibfnamefont{O.~V.} \bibnamefont{Billoni}},
  \bibnamefont{and} \bibinfo{author}{\bibfnamefont{J.~I.}
  \bibnamefont{Perotti}}, \bibinfo{journal}{Physical Review E}
  \textbf{\bibinfo{volume}{101}}, \bibinfo{pages}{012306}
  (\bibinfo{year}{2020}).

\bibitem[{\citenamefont{Zhu and Chen}(2017)}]{zhu2017finite}
\bibinfo{author}{\bibfnamefont{Y.}~\bibnamefont{Zhu}} \bibnamefont{and}
  \bibinfo{author}{\bibfnamefont{X.}~\bibnamefont{Chen}},
  \bibinfo{journal}{arXiv preprint arXiv:1710.02957}  (\bibinfo{year}{2017}).

\bibitem[{\citenamefont{Stauffer and Aharony}(1994)}]{Stauffer}
\bibinfo{author}{\bibfnamefont{D.}~\bibnamefont{Stauffer}} \bibnamefont{and}
  \bibinfo{author}{\bibfnamefont{A.}~\bibnamefont{Aharony}},
  \emph{\bibinfo{title}{Introduction To Percolation Theory}}
  (\bibinfo{publisher}{Taylor \& Francis}, \bibinfo{year}{1994}), ISBN
  \bibinfo{isbn}{9781420074796}.

\bibitem[{\citenamefont{Fujiki and Yakubo}(2020)}]{fujiki2020identification}
\bibinfo{author}{\bibfnamefont{Y.}~\bibnamefont{Fujiki}} \bibnamefont{and}
  \bibinfo{author}{\bibfnamefont{K.}~\bibnamefont{Yakubo}},
  \bibinfo{journal}{Physical Review E} \textbf{\bibinfo{volume}{101}},
  \bibinfo{pages}{032308} (\bibinfo{year}{2020}).

\bibitem[{\citenamefont{Mayo et~al.}(2015)\citenamefont{Mayo, Abdelzaher, and
  Ghosh}}]{mayo2015long}
\bibinfo{author}{\bibfnamefont{M.}~\bibnamefont{Mayo}},
  \bibinfo{author}{\bibfnamefont{A.}~\bibnamefont{Abdelzaher}},
  \bibnamefont{and} \bibinfo{author}{\bibfnamefont{P.}~\bibnamefont{Ghosh}},
  \bibinfo{journal}{Computational Social Networks}
  \textbf{\bibinfo{volume}{2}}, \bibinfo{pages}{1} (\bibinfo{year}{2015}).

\bibitem[{\citenamefont{Fujiki et~al.}(2018)\citenamefont{Fujiki, Takaguchi,
  and Yakubo}}]{fujiki2018general}
\bibinfo{author}{\bibfnamefont{Y.}~\bibnamefont{Fujiki}},
  \bibinfo{author}{\bibfnamefont{T.}~\bibnamefont{Takaguchi}},
  \bibnamefont{and} \bibinfo{author}{\bibfnamefont{K.}~\bibnamefont{Yakubo}},
  \bibinfo{journal}{Physical Review E} \textbf{\bibinfo{volume}{97}},
  \bibinfo{pages}{062308} (\bibinfo{year}{2018}).

\end{thebibliography}

\end{document}